\documentclass[aps,pre,superscriptaddress,longbibliography]{revtex4-1}
\usepackage{graphicx,amsmath}
\usepackage[amssymb]{SIunits}
\newcommand{\makered}[1]{#1}
\usepackage{color}

\begin{document}

\title{Columnar structure formation of a dilute suspension\\ of settling spherical particles in a quiescent fluid}

\author{Sander G. Huisman}
\author{Thomas Barois}
\author{Micka\"el Bourgoin}
\affiliation{Univ Lyon, Ens de Lyon, Univ Claude Bernard, CNRS, Laboratoire de Physique, F-69342 Lyon, France}

\author{Agathe Chouippe}
\author{Todor Doychev}
\affiliation{Institute for Hydromechanics, Karlsruhe Institute of Technology, 76131 Karlsruhe, Germany}

\author{Peter Huck}
\affiliation{Univ Lyon, Ens de Lyon, Univ Claude Bernard, CNRS, Laboratoire de Physique, F-69342 Lyon, France}
\author{Carla E. Bello Morales}
\affiliation{Univ Lyon, Ens de Lyon, Univ Claude Bernard, CNRS, Laboratoire de Physique, F-69342 Lyon, France}
\affiliation{CICATA-Qro, Instituto Polit\'ecnico Nacional, 76090 Quer\'etaro, Mexico}

\author{Markus Uhlmann}
\affiliation{Institute for Hydromechanics, Karlsruhe Institute of Technology, 76131 Karlsruhe, Germany}

\author{Romain Volk}
\affiliation{Univ Lyon, Ens de Lyon, Univ Claude Bernard, CNRS, Laboratoire de Physique, F-69342 Lyon, France}

\date{\today}

\begin{abstract} 
The settling of heavy spherical particles in a column of quiescent fluid is investigated. The performed experiments cover a range of Galileo numbers ($110 \leq \text{Ga} \leq 310$) for a fixed density ratio of $\Gamma = \rho_p/\rho_f = 2.5$. In this regime the particles are known to show a variety of motions [M. Jenny, J. Du\v{s}ek, and G. Bouchet, ``Instabilities and transition of a sphere falling or ascending freely in a Newtonian fluid'', Journal of Fluid Mechanics \textbf{508}, 201--239 (2004)]. It is known that the wake undergoes several transitions for increasing $\text{Ga}$ resulting in particle motions that are successively: vertical, oblique, oblique oscillating, and finally chaotic. Not only does this change the trajectory of single, isolated, settling particles, but it also changes the dynamics of a swarm of particles as collective effects become important even for dilute suspensions with volume fraction up to $\Phi_V = \mathcal{O}\left(10^{-3}\right)$, which are investigated in this work. Multi-camera recordings of settling particles are recorded and tracked over time in 3 dimensions. A variety of analysis are performed and show a strong clustering behavior.  The distribution of the cell areas of the Vorono\"i tessellation in the horizontal plane are compared to that of a random distribution of particles and shows clear clustering. Moreover, a negative correlation was found between the Vorono\"i area and the particle velocity; clustered particles fall faster. In addition, the angle between adjacent particles and the vertical is calculated and compared to a homogeneous distribution of particles, clear evidence of vertical alignment of particles is found. The experimental findings are compared to simulations.
\end{abstract}
\maketitle

\section{Introduction}
The settling of a sphere in a quiescent viscous fluid is a longstanding problem, already explored by Newton~\cite{bib:newton} in the XVII$^{\text{th}}$ century. The problem is simple only in appearance. The usual picture of a straight settling trajectory with a constant terminal velocity resulting from the balance between buoyancy and viscous drag is an important but only marginal situation in a much richer landscape of possible settling regimes.

A sufficiently small or slowly settling sphere (\emph{i.e.} in a low Reynolds number approximation, so that the flow around the particle can be approximated as a Stokes flow) with diameter $d$ and density ratio $\Gamma=\rho_p/\rho_f$ ($\rho_p$ is the particle density and $\rho_f$ the density of the fluid) surrounded by a fluid with viscosity $\nu$, will indeed settle along a straight vertical path, reaching a steady terminal vertical velocity $V$, where the linear viscous drag (due to the Stokes flow around the particle) $F_D=\frac{1}{8}C_D(\text{Re}_p) \rho_f\pi d^2  v_s^2$ (with $\text{Re}_p = d V /\nu$ the particulate Reynolds number and $C_D(\text{Re}_p)$ the drag coefficient, which is simply $24/\text{Re}_p$ for a sphere in the limit $\text{Re}_p \ll 1$) is balanced by the buoyancy force, so that $V=(\Gamma-1)gd^2/\makered{(}18\nu\makered{)}$. Several successive scenarios arise as the size or density of the particle increases and the particulate Reynolds number increases.

\paragraph{Finite size effects}
At finite particulate Reynolds number, the flow around a sphere departs from a simple Stokes flow, and eventually develops wake instabilities. Such instabilities have been extensively studied (numerically and experimentally) in the past 15 years for the case of fixed spheres in a flow~\cite{johnson1999,ghidersa2000,schouveiler2002,bouchet2006}. The global picture is now clear, with several successive instabilities taking place as $\text{Re}_p$ increases: from a steady axisymmetric wake at low $\text{Re}_p$ to, first\makered{,} a wake with steady planar double-threaded vortices (above $\text{Re}_p \simeq 210$), then an unsteady time-periodic vortex shedding wake (above $\text{Re}_p \simeq 275$) and finally a transition towards a fully three dimensional chaotic wake above $\text{Re}_p \simeq 360$). These successive modifications of the wake, have two main implications. First, the value of the drag coefficient $C_D$ departs from the simple $24/\text{Re}_p$ law as $\text{Re}_p$ increases, reaching eventually to a full non-linear drag regime (with an almost constant value of $C_D$ at very high values of $\text{Re}_p$ when the wake becomes fully turbulent). 

\begin{figure}[ht]
	\begin{center}
		\includegraphics[width=0.65\textwidth]{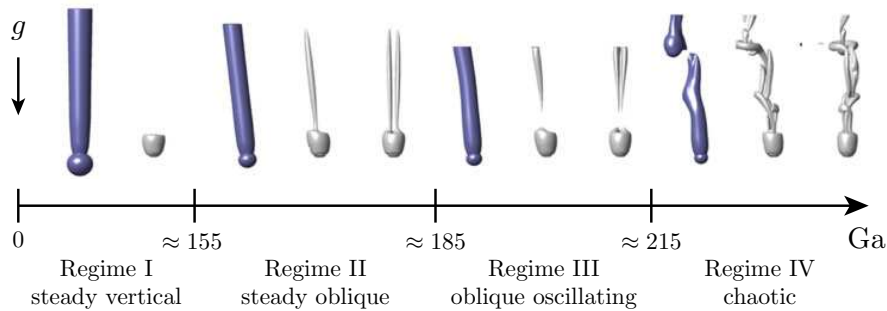}
		\caption{(color online) Four regimes of single particle settling in ambient fluid  as a function of $\text{Ga}$ for $\Gamma=1.5$. The snapshots are for (from left to right) $\text{Ga} = 144$, 178, 190, and 250. The visualizations are taken from \cite{uhlmann14a} and show (in purple) the iso-contour for which the velocity is $1.2V_g$ and show (in gray) from multiple angles the iso-contour of a $\lambda_2$ criterion.}
		\label{fig:regimes} 
	\end{center}
\end{figure}

Second, in the case of a free\makered{-}falling sphere (with additional degrees of freedom compared to the case of a fixed sphere) these wake instabilities induce particle path instabilities marking a departure from a simple vertical trajectory. The settling regime of a free falling sphere is controlled by two dimensionless parameters: 
\makered{i.e.}\ the density ratio $\Gamma=\rho_p/\rho_f$ and the Galileo number $\text{Ga}=V_g d/\nu$, which can be thought of as a Reynolds number based on the gravitational velocity $V_g=[(\Gamma-1)gd]^{1/2}$. A sphere settling in finite $\text{Ga}$ regimes (hence beyond the Stokes approximation) \makered{is} referred to as \makered{a} `finite size sphere' (by definition the settling problem concerns the case $\Gamma > 1$). The settling Reynolds number $\text{Re}_p = V d/\nu$ is then an output parameter of the problem. The path instabilities at finite values of $\text{Ga}$ are now well characterized and have been deeply investigated numerically~\cite{jenny2004} and experimentally~\cite{veldhuis2007_IJMF,horowitz2010_JFM} in the past decade for single finite size free falling spheres, cf.\ the visualizations in fig.~\ref{fig:regimes} (see also the review by Ern \emph{et al.}~\cite{ern2012}). 

\paragraph{Collective effects}
The case of many particles settling simultaneously raises further complexity, as long-range multi-particle hydrodynamic interactions emerge. In the limit of point particles (or equivalently in the limit $\text{Ga}\ll 1$), the collective settling of such an ensemble of spheres and the underlying hydrodynamic interactions can be efficiently studied using Stokesian dynamics methods~\cite{brady1988,ramaswamy2001}. This approach yields a satisfactory quantitative comparison with experiments~\cite{guazzelli2011}, although several questions remain, in particular, regarding the induced fluctuations and the correlation lengths of particles and flow motion. Much less is known however for the case of a settling ensemble of finite size particles. Systematic experimental studies remain scarce. Parthasarathy and Faeth~\cite{parthasarathy1990} and Mizukami et al.~\cite{mizukami1992} have performed a series of experiments in dilute conditions (volume fraction of particles $\Phi_v < 10^{-4}$) in a range of $\text{Ga}$ from 40 to 340. Their experiments focused on measuring the fluctuations of the flow induced by the interacting wakes of the particles and showed that in dilute regimes a linear superposition of wakes gives a good approximation. It is only recently that accurate numerical simulations of a large number of fully resolved finite size particles settling collectively and fully coupled with the fluid have become possible~\cite{uhlmann2005_JCP,kajishima2002}, mainly thanks to immersed boundary methods combined to direct numerical simulations of the Navier-Stokes equation. These methods permit a description of the coupling between the particle and the surrounding flow at the interface level. Uhlmann \& Doychev~\cite{uhlmann14a,doychev14} reported recently using this method that depending on the value of $\text{Ga}$, particles may tend (or not) to align in columnar clusters along their wakes, resulting in an enhancement of their settling velocity (the average settling velocity was found up to 12\% faster than for individual particles). Such a clustering and settling enhancement was observed for $\text{Ga}=178$ (hence in a regime of steady oblique motion for individual settling particles, see fig.~\ref{fig:regimes}), but was not observed for $\text{Ga}=121$ (when individual particles settle along steady vertical path). This study shows that the interplay of individual wake instabilities and collective interactions is crucial to understand the settling of an ensemble of finite size spheres, even in relatively dilute conditions (in their study $\Phi_v=5\makered{\times} 10^{-3}$). However, the computational cost of these simulations does not yet allow a systematic exploration of the parameter space (density ratio, Galileo number\makered{,} and volume fraction). At the moment these studies~\cite{uhlmann14a,doychev14} only comprise three parameter points at a single value of the density ratio and for two different solid volume fractions and Galileo numbers. We propose here to explore the collective settling of finite size spheres experimentally, with the goal to broaden the range of parameters, in particular regarding the role of the Galileo number around the first wake instabilities. In the present article we therefore address the experimental counterpart of the aforementioned simulations, by exploring the settling behavior of a swarm of dense finite size particles. We look more particularly at the eventual emergence of clustering and columnar alignment, and its impact on local and global settling velocity, for increasing values of the Galileo number in the range $\text{Ga}\in[110,310]$ and comparable seeding densities of the order of $\Phi_V \approx 5 \times 10^{-4}$. \makered{We compare our experimental data with an existing numerical simulation \cite{doychev14} with $\text{Ga}=178$, in the steady oblique regime when considering an isolated particle, and at a solid volume fraction $\Phi_V=4.8 \times 10^{-4}$ of the same order in the experiments. The numerical method, the grid resolution, and configuration are identical to case M178 of Uhlmann \& Doychev~\cite{uhlmann14a}. The solid volume fraction, however, was reduced by a factor of ten: it employs the immersed boundary method of Uhlmann~\cite{uhlmann2005_JCP} on a triply periodic box elongated in the direction of gravity with a treatment of collisions with a repulsive force~\cite{glowinski1999}. The simulation domain has an extension of 85 particle diameters in the horizontal directions and 171 diameters in the vertical direction with a uniform grid resolution $\Delta x$ such that $D/\Delta x=24$. This simulation covers 1756 gravitational time units.}

\section{Experimental details}
\begin{figure}[ht]
	\begin{center}
		\includegraphics[width=0.55\textwidth]{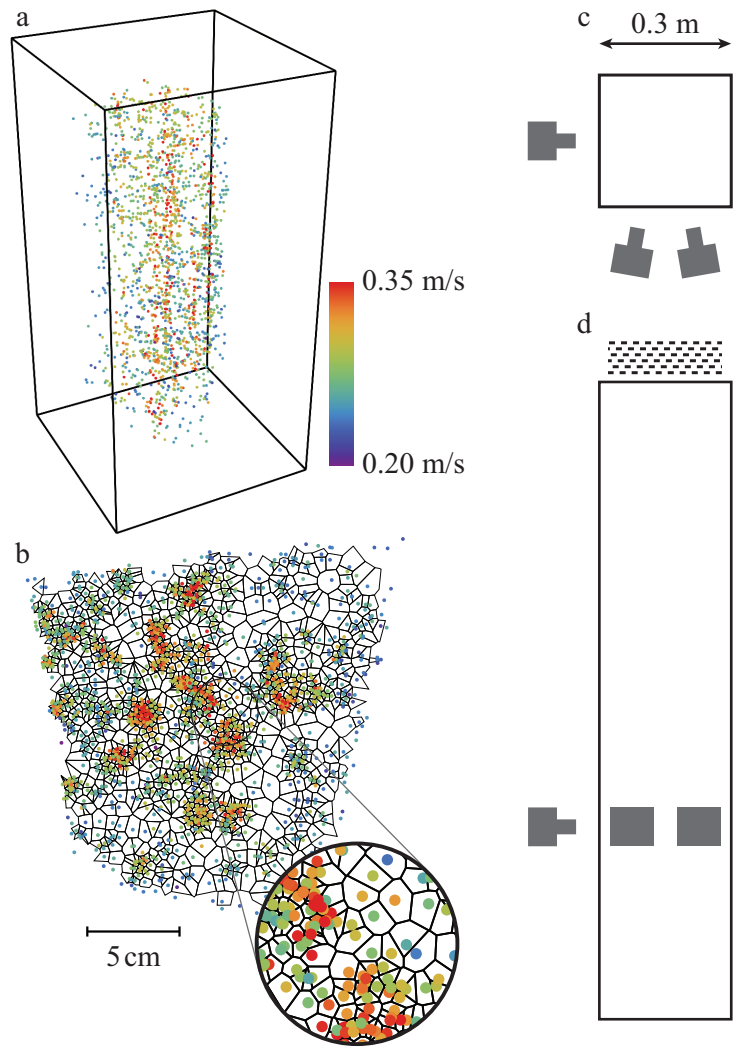}
		\caption{(color online) (a) 3D reconstructing of settling particles with $\text{Ga}=170$ colored by their velocities. Short sides of the box are \unit{0.3}{\meter}. (b) Top view of fig.~a including scale bar and the \makered{corresponding} Vorono\"i tessellation, see also the detailed ($3\times$) zoom. (c) Top view with the camera arrangement. (d) Side view with the injector visible at the top. The injector comprises stacked perforated metal meshes in order to randomly inject the particles.}
		\label{fig:setup}
	\end{center}
\end{figure}
In the current work we look at the collective effects of heavy spherical particles settling in a column of quiescent liquid, see figures \ref{fig:setup}c and \ref{fig:setup}d. The \unit{2}{\meter} high column has a square cross section with sides of \unit{0.3}{\meter}. In order to vary $\text{Ga}$ two sizes of glass particles ($d=\unit{2}{\milli \meter}$ and $d=\unit{3}{\milli \meter}$) with density $\rho_p = \unit{2.5\times10^3}{\kilo \gram \per \meter \cubed}$ and two fluid viscosities ($\nu = \unit{2.05\makered{\times 10^{-6}}}{\makered{\meter^2 \per \second}}$ and $\nu = \unit{3.15\makered{\times 10^{-6}	}}{\makered{\meter^2 \per \second}}$ created using a mixture of water and Ucon$^\mathrm{TM}$ oil) are used, resulting in $\text{Ga} \in [110,310]$. Constant seeding of heavy particles is accomplished by gradually pouring particles on a stack of \makered{6} perforated meshes\makered{, which has become the de facto standard, see e.g. ref.~\cite{parthasarathy1990}.} \makered{The particles that leave the last mesh were checked to have a Vorono\"i area distribution closely resembling an RPP distribution (and very far from the Vorono\"i distribution farther downstream) by injecting particles in a very shallow tank and then taking photographs of the injected particles. The columns are therefore not introduced because of the mesh but are formed during descent.} The particles travel roughly one meter before entering the field of view of the cameras, attaining their terminal velocity far before entering the view. For our largest particles this distance of one meter corresponds to roughly $330d$. Numerical simulations~\cite{uhlmann14a} suggest that at least a distance of $250d$ is needed in order to observe clusters and columns~\cite{uhlmann14a}. The falling particles are recorded by a set of three Flare 2M360-CL cameras from IO Industries at up to \unit{240}{fps} at a resolution of $2048\times 1088$ using a \unit{8}{\milli \meter} focal length lens covering a height of roughly \unit{0.6}{\meter}. The particles are coated black and backlight illumination is employed for enhanced contrast. The cameras are calibrated using an \textit{in-situ} calibration method \cite{machicoane2016} achieving a sub-radius resolution of roughly \unit{300}{\micro \meter}. Picked from a set of experiments are those for which the volume fraction $\Phi_V$ is constant and comparable to one another (see table \ref{tab:experimentoverview}). \makered{Note that the injection is done manually and the volume concentration is only known a posteriori. Injections that are too short do not form columns and only show transient behavior and are therefore rejected. Furthermore we look for experiments where the number of particles in the field of view remains roughly constant. The selected experiments all have a constant volume fraction in the range $10^{-4}$--$10^{-3}$ and have sufficiently long duration of injection such that columns can potentially be formed.} In particular, we make sure to omit the transient part of each experiment, at the beginning when particles start entering the measurement volume and at the end when they leave it. These \makered{experiments} have the following Galileo numbers: $\text{Ga}=110 (2)$, $\text{Ga}=170 (2)$, $\text{Ga}=200 (2)$, and $\text{Ga}=310 (3)$\makered{, h}ere the values between parenthesis are the number of experiments\makered{, see also table \ref{tab:experimentoverview}}. Note that the values for $\text{Ga}$ are rounded to the nearest 10 to reflect the errors in $\rho_f$, $d$, and $\nu$, which lead to an estimated error of 10 for the Galileo number. In the recorded imagery particles are detected and using standard Particle Tracking Velocimetry (PTV) the location of the particles are tracked over time and three dimensional space. A snapshot of the reconstructed particle positions can be seen in fig.~\ref{fig:setup}a for the case $\text{Ga}=170$. Visual inspection reveals already clear vertical trails of particles.

\begin{table}
  \centering
  \begin{tabular}{cccccccccc}
	\hline
      \text{Ga} & $\Gamma$ & $d$ [mm] & $\nu$ [\unit{10^{-6}}{\makered{\meter^2 \per \second}}]  & $V_g$ [m/s] & $V_\infty$ [m/s] & \makered{$\langle V \rangle$ [m/s]} & \makered{$\langle V \rangle$/$V_\infty$} & $\text{Re}_\infty$ & $\phi_V$  \makered{[$10^{-5}$]}\\
	\hline
 110 & 2.5 & 2.0 & 3.15 & 0.17 & 0.18 & \makered{0.22} & \makered{1.19} & 120 & 38 \\
 110 & 2.5 & 2.0 & 3.15 & 0.17 & 0.18 & \makered{0.21} & \makered{1.13} & 120 & 20 \\
 170 & 2.5 & 2.0 & 2.05 & 0.17 & 0.25 & \makered{0.31} & \makered{1.25} & 240 & 49 \\
 170 & 2.5 & 2.0 & 2.05 & 0.17 & 0.25 & \makered{0.30} & \makered{1.21} & 240 & 56 \\
 200 & 2.5 & 3.0 & 3.15 & 0.21 & 0.30 & \makered{0.36} & \makered{1.17} & 290 & 79 \\
 200 & 2.5 & 3.0 & 3.15 & 0.21 & 0.30 & \makered{0.38} & \makered{1.23} & 290 & 100 \\
 310 & 2.5 & 3.0 & 2.05 & 0.21 & 0.33 & \makered{0.37} & \makered{1.10} & 490 & 84 \\
 310 & 2.5 & 3.0 & 2.05 & 0.21 & 0.33 & \makered{0.39} & \makered{1.16} & 490 & 79 \\
 310 & 2.5 & 3.0 & 2.05 & 0.21 & 0.33 & \makered{0.39} & \makered{1.17} & 490 & 76 \\
	\hline
  \end{tabular}
  \caption{Overview of the experimental parameters. $\text{Re}_\infty$ is defined as $\text{Re}_\infty = V_\infty d /\nu$ and $V_g = [(\Gamma - 1) g d]^{1/2}$.}
  \label{tab:experimentoverview}
\end{table} 

\section{Results}
\begin{figure}[ht]
	\begin{center}
		\includegraphics{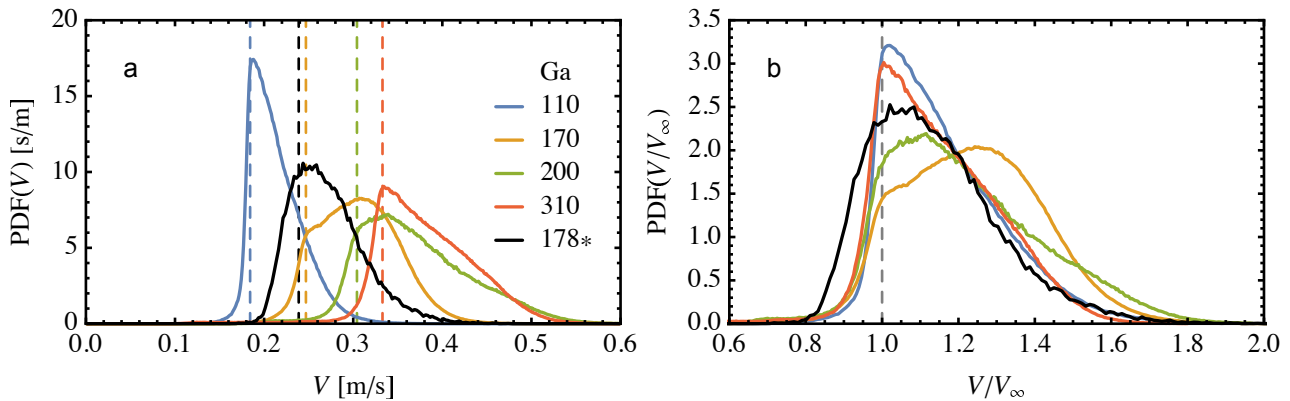}
		\caption{(color online) (a) Velocity probability density function as a function of $\text{Ga}$. Multiple experiments for the same $\text{Ga}$ are in agreement and are combined to improve the quality of the statistics. Dashed lines indicate the settling velocity of isolated particles $V_\infty$ for each $\text{Ga}$ and are colored analogously. Starred data is from numerical simulations \cite{doychev14}, the data is made dimensional using the diameter and viscosity from the experimental $\text{Ga}=170$ case. Note that $\Gamma=1.5$ for the numerical data~\cite{uhlmann14a}, while the experiments have $\Gamma = 2.5$. (b) Vertical velocity normalized by the settling velocity of an isolated particle, see table \ref{tab:experimentoverview}. Same colors as in figure (a). \makered{Average velocities for each experiment can be found in Table \ref{tab:experimentoverview}.}}
		\label{fig:pdfvga}
	\end{center}
\end{figure}

The velocity of the particles is inferred from the particle trajectories, and is defined as positive in the direction of gravity, see also fig.~\ref{fig:setup}a. \makered{Visually we already observe the formation of high-speed (orange--red) columns inside the measurement volume. Moreover, during the experiment one can clearly see the column extend way beyond measurement section. We estimate that the columns are at least \unit{1.5}{\meter} long without any sign that they are unstable or that they break up.} The probability density function (PDF) of the velocity as a function of $\text{Ga}$ can be found in figure \ref{fig:pdfvga}a. The mean settling velocity of isolated particles are included as colored dashed lines and increases with $\text{Ga}$. We complement our experimental data with numerical simulations~\cite{doychev14} with comparable $\text{Ga}=\makered{178}$, $\Gamma = 1.5$, and $\Phi_V = 0.00048$, see fig.~\ref{fig:pdfvga}a. A large spread of the velocity above its isolated velocity is found, which supports the idea of an enhanced settling velocity in a suspension of particles. Figure \ref{fig:pdfvga}b shows the velocities normalized by their isolated settling velocity. We note that the largest velocity enhancement is seen for $\text{Ga}=170$. The speed enhancement for $\text{Ga}=310$ is notably less than for the $\text{Ga}=170$ and $\text{Ga}=200$ cases\makered{, see also table \ref{tab:experimentoverview}}. Several possible reasons might explain this behavior. First, the most probable explanation is that the chaotic wake\makered{s} of the $\text{Ga}=310$ particles \makered{might} prohibit the formation of a stable train of settling particles, which is not the case for intermediate $\text{Ga}$. Second, the volume fraction necessary for `equivalent' clustering might be a function of $\text{Ga}$. The velocities lower than the mean isolated velocity (below $1$ in the fig.~\ref{fig:pdfvga}b) is caused by the combined effects of non-monodispersity, minute density-differences, lack of roundness, and the unavoidable adherence of micro bubbles. In addition, it is clear that particles drag fluid downwards with them, and therefore, when fluid is going downward, and equal amount has to go up as per the continuity equation. This upward flow could cause isolated particles (particles in low density regions) to appear to have a lower velocity in the frame of reference of the lab. This effect can not be disentangled by us as we do not have access to the velocity of the fluid. The experiments are performed in the center of a square box and any upward flow can thus go `around' the falling particles \textit{en masse}, however in the simulation there is no such possibility as the simulation domain is performed in a 3 dimensional periodic box. In the simulation we therefore expect more particles with a velocity below $V/V_\infty = 1$. Indeed, this can clearly be seen in fig.~\ref{fig:pdfvga}b. The strong enhancement of the velocity can be attributed to particles entering the wake of upstream (leading) particles, similar to what happens with aligned bubbles rising in tandem~\cite{huisman2012}. We therefore expect that particles tend to align vertically such that a collection of vertically clustered particles can fall faster as they would individually.

\begin{figure}[ht]
	\begin{center}
		\includegraphics{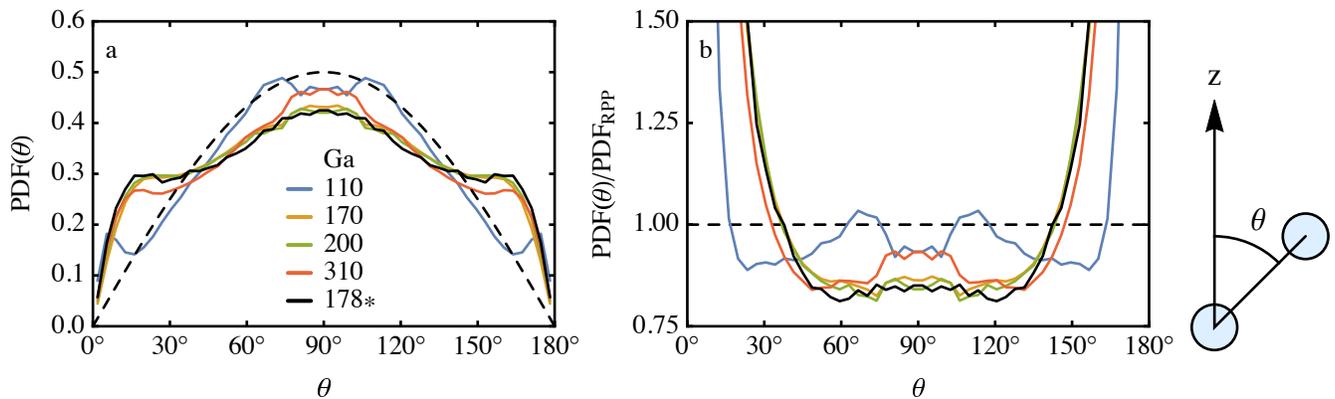}
		\caption{(color online) \makered{a.}~Angular pair probability density function for various $\text{Ga}$. The angle $\theta$ is defined as the angle between two adjacent particles and the vertical, as seen on the right. The dashed lines indicate the PDF for a set of Poisson particles (RPP), \textit{i.e.} non-interacting particles that are place randomly, and is given by $\text{PDF}_\text{RPP} = \sin(\theta)/2$ and is shown for reference. The graph is, by definition, symmetric around $\theta = 90^{\circ}$. \makered{b.~The data of fig.~a normalized by $\text{PDF}_\text{RPP}$. Same colors as in fig.~a. Data bigger than 1 is more likely to occur, while smaller than 1 is less likely to occur as compared to RPP. Strong enhancement can be seen around $\theta=0$ and $\theta=180^{\circ}$.}}
		\label{fig:pdfθga}
	\end{center}
\end{figure}

To investigate this vertical alignment hypothesis we look at the angle $\theta$ of a particle with other particles in its vicinity and the vertical, see the sketch in fig\makered{s}.~\ref{fig:pdfθga}. The focus is at close-range interactions, and therefore the distance between two neighboring particles is limited to $8d$. We plot the PDF of the angle $\theta$ as a function of $\text{Ga}$, see fig.~\ref{fig:pdfθga}\makered{a}.

To compare our data we introduce the concept of a random set of Poisson point (RPP) particles. These non-interacting particles are independently and randomly placed inside a volume of choice. We will use this as reference and show that our particles (despite the low $\text{Re}$ and relatively low $\Phi_V$) do interact with each other as their statistics strongly differ from that of RPP particles. For such a random set of points the distribution of the angle $\theta$ can be theoretically calculated and follows $\text{PDF}_\text{RPP} = \sin(\theta)/2$ where $\theta \in [0,\pi]$, following from the Jacobian of spherical coordinates. Each pair is considered twice and with a different angle $\theta$. The sum of these two angles is however $\unit{180}{\degree}$, which ensures that the PDF is indeed symmetric around $\theta = \unit{90}{\degree}$. For the low $\text{Ga}$ case of $\text{Ga}=110$ we see that the PDF \makered{(fig.~\ref{fig:pdfθga}a)} closely resembles the one that is found if one were to take random particles. We do find a slight increase around $\theta = \unit{0}{\degree}$ and a slight decrease around $\theta = \unit{90}{\degree}$ which might be caused by the mild clustering that we see from visual inspection, and which we will show later using Vorono\"i analysis. For the high $\text{Ga}$ cases ($\text{Ga}\geq170$), we see a very different behavior; namely that for $\unit{45}{\degree}<\theta<\unit{135}{\degree}$ we see a clear reduction in the number of neighbors\makered{, see also fig.~\ref{fig:pdfθga}b where we present the distribution normalized by the RPP distribution.}. So the chance of finding a particle next to each other is significantly reduced for the high $\text{Ga}$ cases. As a consequence, or rather a cause, particles are found to align vertically far more than a random set of Poisson particles; the PDF is found to be much higher for $\theta < \unit{45}{\degree}$ (or equivalently $\theta > \unit{135}{\degree})$\makered{, which can be more clearly seen in figure \ref{fig:pdfθga}b.} Figure\makered{s} \ref{fig:pdfθga} \makered{have} the inclusion of numerical data~\cite{doychev14} with $\text{Ga}=\makered{178}$, $\Phi_V = 0.00048$, and $\rho_p/\rho_f = 1.5$ which is comparable to our set of parameters. For this data we find that angle-pair distribution is remarkably similar to the experimentally found distributions for $\text{Ga}=170$, despite having a fairly different density ratio $\Gamma = \rho_p/\rho_f$. Indeed a difference in $\Gamma$ can have a profound impact on the stability of the wake as found by refs.~\cite{jenny2004,veldhuis2007_IJMF}, though the influence is expected to be relatively small in the range $1.5<\Gamma<2.5$ for this value of the Galileo number. Note, however, that the phase spaces presented in refs.~\cite{jenny2004,veldhuis2007_IJMF} are for the case of vanishing $\Phi_V$, and these lines might shift significantly for increasing $\Phi_V$ as neighboring particles can trigger wake instabilities.

A look from the top further corroborates the hypothesis that these particles cluster and thereby form a rapidly-settling group (or cluster) of particles, see fig.~\ref{fig:setup}b. This top view shows the horizontal position of the particles colored by their velocities. Not only do we observe a set of high-density and low density-regions---far different from a RPP---but also that the settling velocities are higher in high-density regions, which can be attributed to vertical alignment and further corroborates our view of the mechanism of enhanced settling velocity. The observation of relatively low and high density regions---different from RPP---can be substantiated by computing the Vorono\"i tessellation in the horizontal plane. It is indeed the Vorono\"i tessellation in the horizontal plane that should clearly show a signature of clustering as it is this plane that is perpendicular to gravity. An example of a Vorono\"i tessellation is included in fig.~\ref{fig:setup}b. \makered{We take particles over the entire measurement height (roughly \unit{550}{\milli \meter}) such as to have as many samples as possible. Note that the choice of the vertical extent needs to be at least several times the typical vertical distance between descending particles in the same column, such that vertically aligned particles create dense regions once projected on the horizontal plane. We have checked that the PDF of the Vorono\"i areas is similar if a limited vertical extent (down to \unit{125}{\milli \meter}) is chosen.} Each particle is assigned the set of all points (forming a convex polygon called a Vorono\"i cell) that is closest to itself rather than any other particle. The area of the cell is now inversely proportional to the local density~\cite{monchaux2010}. Mathematically, the Vorono\"i tessellation is the dual graph of the Delaunay triangulation. We make sure not to include Vorono\"i cells at the periphery, which create artificially large cells, skewing the statistics.

\begin{figure}[ht]
	\begin{center}
		\includegraphics{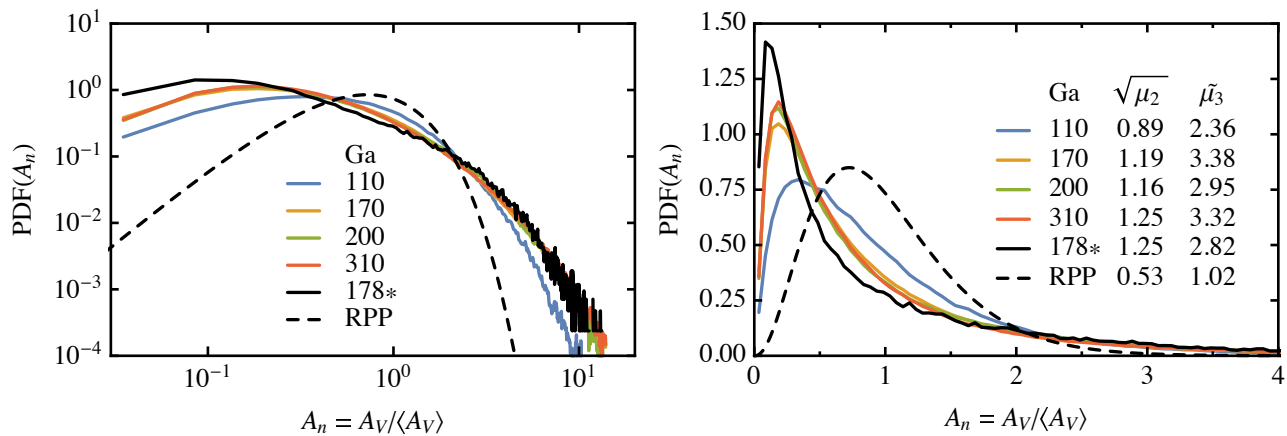}
		\caption{(color online) Probability distribution of the normalized Vorono\"i areas in the horizontal plane. The corresponding probability density function for randomly distributed points (RPP) is shown as a dashed line. Left: Logarithmic axes. Right: Linear axes, the legend includes the standard deviation ($\sqrt{\mu_2}$) and the skewness ($\tilde{\mu_3}$) of $A_n$.}
		\label{fig:voronoiareas}
	\end{center}
\end{figure}

The Vorono\"i tessellation is calculated for each frame and the area of the cells ($A_\text{V}$) are scaled by the mean area \makered{of each frame}. The PDF of the normalized area ($A_n = A_V/\left\langle A_V \right\rangle$) is calculated for each $\text{Ga}$, see figs.~\ref{fig:voronoiareas}. The figures include the corresponding PDF for random Poisson particles (RPP). For the lowest value of the Galileo number, $\text{Ga}=110$ we already observe an increase in probability density for smaller and larger cells indicating some clustering, as we also observe in the angle-pair distribution in fig\makered{s}.~\ref{fig:pdfθga}. For large $\text{Ga}$, we find even more increase in the probability density for both small and large Vorono\"i areas, indicating a more pronounced clustering. Such an important level of clustering could already be observed from a simple visual inspection of fig.~\ref{fig:setup}b: more dense and `open' areas than one would expect from randomly placed particles. Broadening of the distribution is also quantified by calculating the standard deviation of the distribution, see figs.~\ref{fig:voronoiareas}. Compared to an RPP ($\sqrt{\mu_2}=0.53$) we indeed observe an increased standard deviation for $\text{Ga}=110$ ($\sqrt{\mu_2}=0.89$), indicating clustering, and further increased standard deviations for even higher $\text{Ga}\geq 170$ (up to $\sqrt{\mu_2}=1.25$). Increased standard deviation with respect to an RPP is indeed also what one observes in fig.~\ref{fig:setup}b. \makered{We emphasize that the particles follow an RPP distribution when they exit the injector---very different from the distribution farther downstream.}

\begin{figure}[ht]
	\begin{center}
		\includegraphics{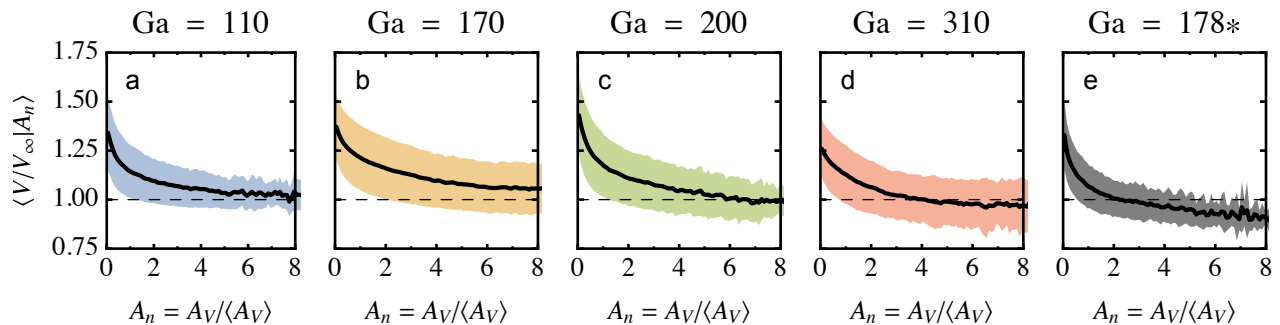}
		\caption{(color online) Velocity of a particle conditioned on its normalized Vorono\"i area in the horizontal plane. Bands represent $\pm \sigma(v|A_n)_{A_n}$. Velocities are normalized using $V_\infty$; the terminal velocity of an isolated particle settling in a quiescent fluid. (e) Numerical data from \cite{doychev14}. Note that the `spiky' data for high $A_n$ are caused by lack of statistics, and that the experiments have more statistics than the very expensive numerical simulations.}
		\label{fig:vaconditioned}
	\end{center} 
\end{figure}

A visual inspection of fig.~\ref{fig:setup}b also suggests that higher velocities are attained at high-density regions. To substantiate that claim the conditional average of the average settling velocity is calculated, see figs.~\ref{fig:vaconditioned}. The average velocity is conditioned on the normalized Vorono\"i area $A_n$ which is proportional to the inverse local density. Moreover, along with the mean velocity also the standard deviation is calculated for each $A_n$. The mean velocities and their spread are normalized using the terminal velocity of an isolated particle settling in a quiescent fluid bath. A clear increase in speed of up to \unit{40}{\%} can be seen for high density (low $A_n$) regions for all $\text{Ga}$, and the graphs are very similar, showing the same behavior. As we stated before, the simulations of ref.~\cite{uhlmann14a} are performed in a 3 dimensional periodic box where upward flow will certainly affect the velocity of some isolated particles. We can now clearly see this in fig.~\ref{fig:vaconditioned}e; isolated particles ($A_n\gg 1$) have normalized velocities below 1, meaning they settle slower than they would if they were to settle by themselves in an infinite bath. Also here we see that the case of $\text{Ga}=310$ shows slightly less enhanced velocity (for example at $A_n=0$) as compared to the $\text{Ga}=170$ and $\text{Ga}=200$ cases. 

\begin{figure}[ht]
	\begin{center}
		\includegraphics{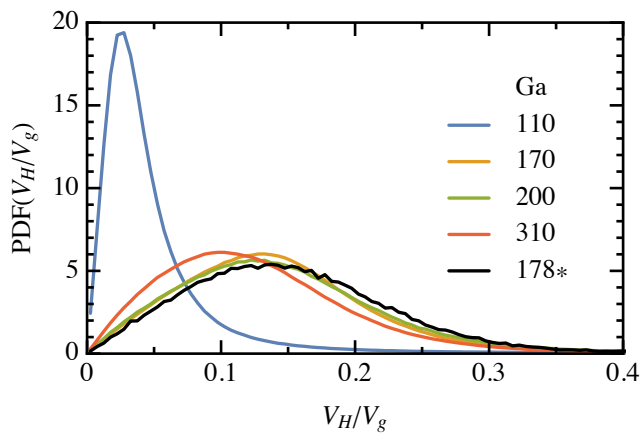}
		\caption{\makered{(color online) PDF of the magnitude of the horizontal velocity ($V_H = \sqrt{V_x^2+V_y^2}$) normalised by the gravitational velocity. Multiple experiments for the same $\text{Ga}$ are in agreement and are combined to improve the quality of the statistics. Starred data is from numerical simulations and has $\Gamma=1.5$~\cite{doychev14,uhlmann14a}, while the experiments have $\Gamma = 2.5$.}}
		\label{fig:vh}
	\end{center}
\end{figure}

\section{Conclusion}
Starting from the visual observation in fig.~\ref{fig:setup}a that particles look vertically aligned, we find that, indeed, particles are preferentially aligned vertically as per our findings of the PDF of $\theta$ in fig\makered{s}.~\ref{fig:pdfθga}. In the two dimensional top view visual inspection suggests low and high density regions with low and high velocities, respectively. We confirm this view in figures \ref{fig:voronoiareas} and \ref{fig:vaconditioned}. Altogether, our findings indicate that particles tend to settle in a preferential columnar configuration, with an increased trend to alignment with increasing Galileo number. This columnar alignment then impacts the settling of the particles as particles that follow in the wake of another particle tend to settle faster. It was found in previous numerical simulations at $\text{Ga}=121$ and $\text{Ga}=178$ (although at \makered{a} higher seeding density $\Phi_V=5\times 10^{-3}$) by Uhlmann \& Doychev\cite{uhlmann14a} that only the latter tended to form columns impacting the settling. The different behavior between the low and large Galileo cases is interpreted in terms of the properties of particle wakes. The wake\makered{s} of the low $\text{Ga}$ particles ($\text{Ga}\leq 155$) have a stable vertical wake (see fig.~\ref{fig:regimes}) and particle fall straight down. Only when two particles are already vertically aligned will the trailing particle fall in the wake of the leading particle. For higher $\text{Ga}$ the wake\makered{s} of the particles are unstable, causing the trajectory of said particles to be either steady oblique ($\text{Ga}\geq 155$), oblique oscillating ($\text{Ga} \geq 185$), or even chaotic ($\text{Ga}\geq 215$). In the aforementioned simulations, the case $\text{Ga}=178$, was in a steady oblique regime, particles therefore have a horizontal motion with a higher chance of `catching' the slipstream of an upstream particle \cite{uhlmann14a}, causing the columnar alignment and the enhanced settling velocity. Our experimental results show that the same behavior (columnar alignment and enhanced settling) prevails for the oblique oscillating ($\text{Ga}=200$) and chaotic ($\text{Ga}=310$) situations which all exhibit lateral motion. The strongest impact on overall settling velocity enhancement, is observed for particles at intermediate Galileo number ($\text{Ga}=170$) in the steady oblique situation. The highest Galileo case explored ($\text{Ga}=310$) was found on the contrary to exhibit less enhanced settling. \makered{At first glance this may seem consistent with the slightly reduced alignment observed for the $\text{Ga}=310$ case from the angular statistics shown in fig.~\ref{fig:pdfθga}. However Vorono\"i statistics in fig.~\ref{fig:voronoiareas} seem on the other hand to exhibit a slightly enhanced clustering for this same case, which would be on the contrary associated to settling enhancement. A first attempt of interpretation could be related to the chaotic nature of the particle motion at $\text{Ga}=310$ which could still promote the chance for particles to catch each others wake and form columns, while disturbing the alignment within the columns. However, the statistics of the horizontal velocity of the particles reveal a more complex situation. The PDF of the amplitude of the horizontal velocity $V_H = \sqrt{V_x^2+V_y^2}$ of the particles for $\text{Ga}=310$ is indeed only marginally different  as compared to the $\text{Ga}=170$, $\text{Ga}=178$, and $\text{Ga}=200$ cases, but very different from the $110$ case, see fig.~\ref{fig:vh}. The $\text{Ga}=110$ case exhibits a low (mean) sideways velocity compared to the other cases, with a narrow distribution, as expected for particles settling mostly in a `steady-straight' regime. The other cases exhibit all very similar PDFs compared to each other, with larger mean sideways velocities and broader distributions (note, though, that the mean sideways horizontal velocity is a little less for the $\text{Ga}=310$ case as compared to the $\text{Ga}=170$, $\text{Ga}=178$, and $\text{Ga}=200$ cases). Purely oblique trajectories (as expected for individual particles settling at $\text{Ga}=170$) would exhibit a larger mean sideway motion compared to $\text{Ga}=110$, but still a narrow PDF (the amplitude of the horizontal velocity being mostly constant). The chaotic case on the contrary is expected to have a wide PDF for the horizontal velocity which undergoes erratic fluctuations. The similarity of the PDFs for the four cases $\text{Ga}=\{170,178,200,310\}$ therefore indicates that, though these particles follow different paths (fig.~\ref{fig:regimes}) when they settle individually, an ensemble of particles eventually shows complex trajectories and chaotic wakes, probably due to particles and the wakes interacting with each other, causing horizontal velocities of the particles to have important and similar fluctuations (see fig.~\ref{fig:vh}). Furthermore, note that fig.~\ref{fig:vh} shows the horizontal velocity of the particles and that the horizontal velocities of the flow in the wake of individual particles can follow a different trend. Overall these observations show that it is very likely irrelevant to speculate on possible interpretations for the differences in the settling collective behavior for particles at $\text{Ga}=170$ and at $\text{Ga}=310$ simply considering the usual `single particle regimes' in fig.~\ref{fig:regimes}, although these regimes certainly play a role in the initial triggering of column formation. Why the $\text{Ga}=310$ case shows less enhanced velocity, while still having a comparable $\text{PDF}(\theta)$, $\text{PDF}(A_n)$, $\left \langle V/V_\infty|A_n \right \rangle$, and $\text{PDF}(V_H/V_g)$ (though in each case it slightly differs from the $\text{Ga}=170$, $\text{Ga}=178$, and $\text{Ga}=200$ cases), is still an open question. Further research, where also the fluid-velocity is measured, might help to find an explanation.} 

An increase or decrease in settling \makered{velocity} is significant in cases where one wants to predict or prevent settling particles in \textit{e.g.}~chemical reactors with solid reactants or settling of particulate matter in riverbeds, and can probably also affect rain falls (the transition between steady vertical and steady oblique regimes for water droplets in air occurs for droplet diameters of the order of 850 microns) and ash cloud dynamics. 

Finally, a striking observation of the present study is that, contrary to the simulations, we do find in the experiment that particles in the steady vertical regime ($\text{Ga}=110$) also exhibit mild clustering and column formation, although less pronounced than for particles at larger Galileo number (as shown by Vorono\"i statistics in fig.~\ref{fig:voronoiareas} and the angular pair statistics in fig\makered{s}.~\ref{fig:pdfθga}). A possible reason for the emergence of columns for such a low Galileo number, where no wake instability is expected, can be related to the existence of a large scale \makered{flow} caused by a \makered{ensemble} of settling particles in a closed container as ours. Such a large-scale flow is absent in the simulations presented in the manuscript as it has a periodic domain. \makered{This large scale flow might induce some different flow dynamics as particles in low-density regions are less affected by upward flow as compare to those particles for the case of the simulation.}

To summarize, we find a coherent set of observations for settling particles that explain the observed features: trajectory properties, vertical alignment, high density regions, and enhanced settling velocity. Future studies, will explore further the role of increasing the seeding density as well as the importance of confinement and boundary conditions. Another important extension of the present study concerns the impact of surrounding turbulence on the column formation and settling enhancement.

\begin{acknowledgments}
We acknowledge the German-French program procope 57129319, the French program ANR-12-BS09-0011 ``TEC2'', and the German Research Foundation (DFG) under Project UH242/1-2 for funding this study.

\end{acknowledgments}


%

\end{document}